\documentclass[conference]{IEEEtran}
\IEEEoverridecommandlockouts
\usepackage{cite}
\usepackage{amsmath,amssymb,amsfonts}
\usepackage{graphicx}
\usepackage{textcomp}

\allowdisplaybreaks

\usepackage[usenames,dvipsnames]{color}
\usepackage{amssymb}
\usepackage{hyperref}
\usepackage{algorithm}
\usepackage{algpseudocode}

\newenvironment{subequations_opti}{%
  \refstepcounter{equation}%
  \setcounter{parentequation}{\value{equation}}%
  \setcounter{equation}{0}%
  \def\theequation{\theparentequation\alph{equation}}%
  \ignorespaces
}{%
  \setcounter{equation}{\value{parentequation}}%
  \ignorespacesafterend
}

\newenvironment{changetheequation}[1]{%
  \def\theequation{#1}%
  \ignorespacesafterend
}{\ignorespacesafterend}

\begin{document}

\title{Max-Min Fairness for IRS-Assisted Secure Two-Way Communications\\ \vspace{-0.5cm}
\thanks{Harindu Jayarathne, Tharindu Wickremasinghe, Kasun T. Hemachandra and Tharaka Samarasinghe are with the University of Moratuwa, Sri Lanka. Saman Atapattu is with the RMIT University, Australia. (Email: \{170258l, 180701b, kasunh, tharakas\}@uom.lk, saman.atapattu@rmit.edu.au)

}
}


\author{\IEEEauthorblockN{Harindu Jayarathne,
Tharindu Wickremasinghe,
Kasun T. Hemachandra,
Tharaka Samarasinghe,\\ and
Saman Atapattu}
}

\IEEEaftertitletext{\vspace{-1\baselineskip}}

\maketitle

\begin{abstract}
This paper investigates an intelligent reflective surface (IRS) assisted secure multi-user two-way communication system. The aim of this paper is to enhance the physical layer security by optimizing the minimum secrecy-rate among all user-pairs in the presence of a malicious user. The optimization problem is converted into an alternating optimization problem consisting of two sub-problems. Transmit power optimization is handled using a fractional programming method, whereas IRS phase shift optimization is handled with semi-definite programming. The convergence of the proposed algorithm is investigated numerically. The performance gain in minimum secrecy-rate is quantified for four different user configurations in comparison to the baseline scheme. Results indicate a 3.6-fold gain in minimum secrecy rate over the baseline scheme when the IRS is positioned near a legitimate user, even when the malicious user is located close to the same legitimate user.
\end{abstract}


\section{Introduction}

Intelligent reflective surface (IRS) is envisioned a promising technology for improved spectral and energy efficiency in 6G and beyond wireless networks~\cite{wu2019intelligent}. An IRS consists of an array of passive elements that induce predetermined phase shifts on the incident signals, creating a smart and controllable signal propagation environment. IRS assisted physical-layer security (PLS) is an area that recently received attention. In the presence of a malicious user, an IRS can coherently combine the signals at legitimate receivers to enhance the information rate, while suppressing the information leakage to the malicious user. 

Maximization of the achievable secrecy rate for IRS aided single-user multiple-input single-output (MISO) system is discussed in \cite{cui2019secure,yu2019enabling}. This system model is extended in~\cite{yu2020robust,xu2019resource} to address multi-user MISO systems, focusing on maximizing the network-wide sum-secrecy rate. Furthermore, IRS aided secure multiple-input multiple-output (MIMO) systems are considered in~\cite{hong2020artificial,dong2020secure}. In~\cite{chen2019}, minimum secrecy-rate is optimized for a multi-user MISO system consisting of multiple malicious users. 

Under two-way communications, both users exchange information over a shared channel, improving the spectral efficiency compared to one-way communication \cite{atapattu2020}. The authors of \cite{atapattu2020} analyze the IRS assisted two-way communication system and maximize the minimum signal-to-interference-plus-noise ratio (SINR) among the two users ensuring fairness. In \cite{wijewardena2021}, the sum-secrecy rate of a two-way communication system consisting of two legitimate users and a malicious user is optimized, where the non-convex IRS phase-shift optimization is tackled using successive convex approximation (SCA) and semi-definite relaxation. However, the proposed techniques are restricted to a single user pair. The authors of \cite{lv2020secure} analyze the performance limits of a secure multi-pair two-way communication system, which incorporates user scheduling for each user pair. However, in \cite{lv2020secure}, all users including the malicious user, do not share a common channel. Thus, resource allocation for secure multi-user two-way communication over a shared channel remains unresolved.

Motivated by these aspects, this paper investigates an IRS assisted system model consisting of multiple user-pairs engaging in two-way communications, sharing a common channel with a malicious user. To ensure the fairness of the secrecy rate attributable to each user pair, minimum secrecy rate is maximized with respect to the IRS phase-shifts and transmit powers. In this scenario, the interference between the legitimate users impact both the information exchange and the information leakage to the malicious user, where the ability of the IRS to assist the legitimate users is worth exploring. Since the optimization variables are coupled in the objective function, it is difficult to optimize jointly with respect to the both variables. Therefore, the problem is first decomposed into two sub-problems and then solved using a sub-optimal alternating optimization (AO) algorithm. The resulting sub-problems are non-convex due to the difference of two convex functions in the objective. Therefore, the IRS phase-shift optimization problem is tackled through SCA and semi-definite relaxation, whereas the transmit power optimization is handled using a fractional programming (FP) approach. The contributions of this paper are summarized below:
\vspace{-0.1cm}
\begin{enumerate}
\item We propose an iterative algorithm to maximize the minimum secrecy-rate among user-pairs in an IRS assisted two-way communication system with respect to IRS phase shifts and the transmit powers of all legitimate users in the presence of a malicious user. The convergence is numerically analyzed for both IRS phase shift optimization and transmit power optimization, which are considered as two separate sub-problems.

\item The numerical results show substantial gains up to 3.6-fold in minimum secrecy rate, when both the IRS and the malicious user are in close proximity to a legitimate user. The influence of the IRS in four different user configurations are quantified compared to the baseline scheme of random IRS phase shifts.
    
\end{enumerate}

\IEEEpubidadjcol

\section{System Model}
\begin{figure}[!t]
\centering
\includegraphics[width=0.7\columnwidth]{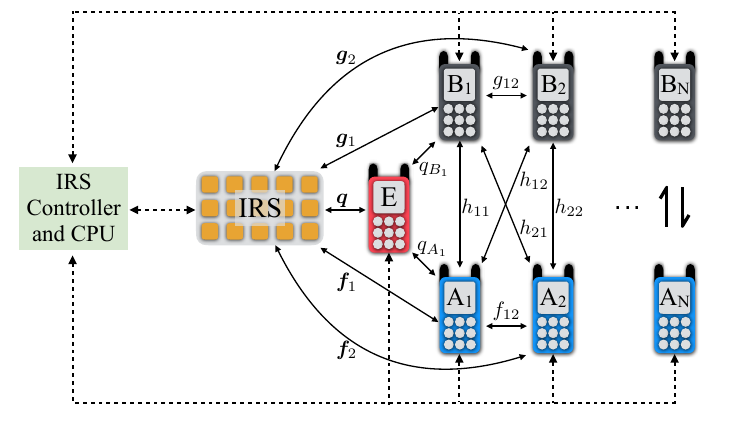}
\caption{IRS assisted secure multi-user two-way communication model.}
\label{fig_system}

\vspace{-0.3cm}
\end{figure}

We consider an IRS assisted two-way communication system as shown in Fig. \ref{fig_system}. The communication environment consists of $N$ pairs of legitimate users denoted by $\mathcal{P} \triangleq \{1,\hdots,N\}$. For $j \in \mathcal{P}$, $A_j$ and $B_j$ denote the two legitimate users belonging to the $j^\text{th}$ user pair. Let $\mathcal{U} \triangleq \{A_1, B_1,\ldots, A_N, B_N\}$ denote the set of all legitimate users in the system. Each user is equipped with a single-antenna receiver and a single-antenna transmitter, facilitating in-band full-duplex (FD) communications. They are exposed to a single-antenna malicious user $E$, that can overhear the confidential information exchanged between the user pairs. An IRS consisting of $L$ elements is deployed to assist the secure communications. It is assumed that the global channel state information (CSI) of all the users are made available at the central processing unit (CPU). The control information and the CSI are exchanged using low bandwidth links, and channel reciprocity is exploited to estimate the CSI.

For $j,k \in \mathcal{P}$, $j\neq k$, let $f_{jk} \in \mathbb{C}^{1\times1}$ denote the channel between $A_j$ and $A_k$, and $g_{jk} \in \mathbb{C}^{1\times1}$ denote the channel between $B_j$ and $B_k$. Let $h_{jk} \in \mathbb{C}^{1\times1}$ denote the channel between $A_j$ and $B_k$, $\boldsymbol{f}_j \in \mathbb{C}^{L\times1}$ the channel between $A_j$ and the IRS, and $\boldsymbol{g}_k \in \mathbb{C}^{L\times1}$ the channel between $B_k$ and the IRS. Moreover, the channels from $A_j$ and $B_k$ to $E$ are denoted by $q_{A_{j}} \in \mathbb{C}^{1\times1}$ and $q_{B_{k}} \in \mathbb{C}^{1\times1}$, respectively. The channel from $E$ to the IRS is denoted by $\boldsymbol{q} \in \mathbb{C}^{L\times1}$. Let $\boldsymbol{\omega}_{ind}^\dag = [e^{j\phi_{1}},\ldots,e^{j\phi_{L}}]$ denote the phase shift matrix of the IRS, where $(.)^\dag$ represents the conjugate transpose. The phase shift of the $i$-th element of the IRS is denoted by $\phi_{i} \in [0,2 \pi)$, $i \in \{ 1,\ldots,L\}$. We have assumed full signal reflection for the ease of practical implementation and to maximize the reflected signal power from the IRS.
The effective channel between $A_j$ and $B_k$ due to the direct and the cascaded user-IRS-user links can be expressed as $\boldsymbol{\omega}^\dag \mathbf{H}_{A_j,B_k}$, where $\boldsymbol{\omega}^\dag = [ \boldsymbol{\omega}_{ind}^\dag \;\; 1 ]$, $\mathbf{H}_{A_j,B_k}^\top =  [\boldsymbol{f}_j^{\top} diag(\boldsymbol{g}_k) \; h_{jk}] \in \mathbb{C}^{1\times (L+1)}$ and matrix transpose is denoted by $(.)^\top$. Similarly, the effective channel between $A_j$ and $E$ can be expressed as $\boldsymbol{\omega}^\dag \mathbf{H}_{A_j,E}$, where $\mathbf{H}_{A_j,E}^\top =  [\boldsymbol{f}_j^{\top} diag(\boldsymbol{q}) \; q_{A_{j}}] \in \mathbb{C}^{1\times (L+1)}$. 


For $j \in \mathcal{P}$, the information symbols of $A_j$ and $B_j$ are denoted as $s_{A_j} \sim \mathcal{CN}(0,1)$ and $s_{B_j} \sim \mathcal{CN}(0,1)$ respectively, with their transmit powers represented as $P_{A_j}$ and $P_{B_j}$. Let $\mathbf{P} = [P_{A_1}, P_{B_1},\ldots, P_{A_N}, P_{B_N}]^{T}$ denote all the transmit powers of legitimate users. The received signal at the user $A_j$ can be expressed as $y_{A_j} =  \sqrt{P_{B_j}} \boldsymbol{\omega}^\dag \mathbf{H}_{B_j,A_j}s_{B_j} + n_{A_j} + l_{A_j} + \sum_{\substack{X\in\mathcal{U} \\ X\neq A_j,B_j}}  \sqrt{P_{X}} \boldsymbol{\omega}^\dag \mathbf{H}_{X,A_j}s_{X}$, where $n_{A_j} \sim \mathcal{N}(0,\sigma^2)$ denotes the additive white Gaussian noise (AWGN) at $A_j$. The residual loop-interference due to FD-operation at $A_j$ is denoted by $l_{A_j}$, which is assumed to be a Gaussian random variable of zero mean and variance $\sigma_{l}^2$. Using the global CSI and $\boldsymbol{\omega}_{ind}^\dag$ obtained from the CPU, the self-interference term $\sqrt{P_{A_j}} \boldsymbol{\omega}^\dag \mathbf{H}_{A_j,A_j}s_{A_j}$ arising from $A_j$'s own signal reflected back from the IRS is assumed to be suppressed. The received signal at $B_j$ can be similarly expressed. Considering that $E$ is eavesdropping the signals transmitted by $j^\text{th}$ user pair, the received signal at $E$ can be arranged as $y_{E} =  \sqrt{P_{A_j}} \boldsymbol{\omega}^\dag \mathbf{H}_{A_j,E}s_{A_j} + \sqrt{P_{B_j}} \boldsymbol{\omega}^\dag \mathbf{H}_{B_j,E}s_{B_j} + n_{E}  +
 \sum_{\substack{X\in\mathcal{U} \\ X\neq A_j,B_j}}  \sqrt{P_{X}}  \boldsymbol{\omega}^\dag \mathbf{H}_{X,E}s_{X}$, where $n_{E} \sim \mathcal{N}(0,\sigma^2)$ is the AWGN noise at $E$. The information rate achieved by $A_j$ is given as
\begin{equation}
\label{eqn_dbl_x}
\begin{split}
& I(y_{A_j};s_{B_j}) \\& = \text{log}_2 \left(\!1+ \frac{P_{B_j}|\boldsymbol{\omega}^\dag \mathbf{H}_{B_j,A_j}|^2}{\sum_{\substack{X\in\mathcal{U} \\ X\neq A_j,B_j}}  P_{X}|\boldsymbol{\omega}^\dag \mathbf{H}_{X,A_j}|^2 + \sigma^2 + \sigma_{l}^2}\right)
\end{split}
\end{equation}
and the information rate $I(y_{B_j};s_{A_j})$ achieved by $B_j$ is defined analogously.
If $E$ attempts to arbitrarily select $j^\text{th}$ user pair and intercept the signals transmitted by $A_j$ and $B_j$, the worst-case information leakage rate at $E$ can be expressed as
\begin{equation}
\label{eqn_dbl_y}
\begin{split}
& I(y_E;s_{A_j},s_{B_j}) \\&= \text{log}_2 \left(1+ \frac{P_{A_j}|\boldsymbol{\omega}^\dag \mathbf{H}_{A_j,E}|^2 + P_{B_j}|\boldsymbol{\omega}^\dag \mathbf{H}_{B_j,E}|^2}{\sum_{\substack{X\in\mathcal{U} \\ X\neq A_j,B_j}}  P_{X}|\boldsymbol{\omega}^\dag \mathbf{H}_{X,E}|^2 + \sigma^2}\right).
\end{split}
\end{equation}
The sum secrecy-rate achieved by $j^\text{th}$ user pair is expressed as
\begin{equation}
\label{eqn3}
\begin{split}
C_{j}(&\mathbf{P},\boldsymbol{\omega}) \!\! = \!\! \left[I(y_{A_j};s_{B_j}\!)\! +\! I(y_{B_j};s_{A_j}\!)\! -\! I(y_E;s_{A_j},s_{B_j}\!) \!\right]^{+}
\end{split}
\end{equation}
where $[x]^{+} = \max \{0,x\}$ \cite{wijewardena2021,wang2012distributed}.

\section{Minimum Secrecy-Rate Maximization}

In this section, we formulate an optimization problem to maximize the minimum secrecy-rate achievable among the user pairs by jointly optimizing the IRS phase shift matrix $ \boldsymbol{\omega} $ and the transmit powers $\mathbf{P}$. We propose an alternating optimization (AO) approach to optimize the objective function iteratively. The optimization problem can be written as
\begin{changetheequation}{P1}
\begin{subequations_opti}\label{eq:6}
\begin{align}
\text{(P1):} \quad \mathop{\max_{\boldsymbol{\omega},\mathbf{P}}} \; \min_{j\in\mathcal{P}} & \quad  C_{j}(\mathbf{P},\boldsymbol{\omega}) \label{eq:6A}\\
\textrm{s.t.} \quad & |\omega^{(i)}| = 1, & \quad & \forall 1 \leq i \leq L, \label{eq:6B}\\
  &\omega^{(L+1)} = 1,   \label{eq:6C}\\ & P_{\text{min}} \leq P_{X} \leq P_{\text{max}}, & \quad & \forall X\in\mathcal{U}, \label{eq:6D}
\end{align}
\end{subequations_opti}
\end{changetheequation}
where the $i$-th element of the vector $\boldsymbol{\omega}$ is denoted by $\omega^{(i)}$. The constraint \eqref{eq:6B} ensures that only the phase of the reflected signal is altered by the IRS elements. The constraint \eqref{eq:6D} guarantees that the transmit power at each user is bounded by maximum and minimum feasible power thresholds $P_{\text{max}}$ and $P_{\text{min}}$, respectively. Due to the non-convex objective function \eqref{eq:6A} and the constraint \eqref{eq:6B}, \eqref{eq:6} is clearly a non-convex problem. The two optimization variables are coupled in the objective function, making the problem jointly non-convex with respect to both variables. Therefore, obtaining an optimal solution with a reasonable level of computational complexity is difficult. Thus, we decouple the problem into two sub-problems and propose an iterative algorithm to solve \eqref{eq:6} efficiently.

\subsection{IRS Phase Shift Optimization}

Initially, \eqref{eq:6} is converted to its equivalent epigraph form \eqref{eq:8} by introducing an auxiliary variable $t$. 
We apply semi-definite relaxation (SDR) by employing the substitution $|\boldsymbol{\omega}^\dag \mathbf{H}_{X,Y}|^2 = \text{Tr}(\mathbf{\tilde{H}}_{X,Y} \boldsymbol{W})$, $X,Y \in \mathcal{U}$, where $\mathbf{\tilde{H}}_{X,Y} = \mathbf{H}_{X,Y} \mathbf{H}_{X,Y}^\dag$, $\boldsymbol{W} = \boldsymbol{\omega} \boldsymbol{\omega}^\dag$ and $\text{Tr}(.)$ denotes the matrix trace. For an arbitrary $\nu \in \mathbb{C}$ satisfying $|\nu| = 1$, define $\tilde{\boldsymbol{\omega}}^\dag = \nu \boldsymbol{\omega}^\dag$. Since $\boldsymbol{W} = \boldsymbol{\omega} \boldsymbol{\omega}^\dag = \tilde{\boldsymbol{\omega}} \tilde{\boldsymbol{\omega}}^\dag$, \eqref{eq:6C} can be modified as $|\omega^{(L+1)}| = 1$ \cite{wijewardena2021}. After substituting and rearranging terms in \eqref{eqn3}, we obtain the updated expression for the user pair specific sum-secrecy rate $G_{j}(\mathbf{P},\boldsymbol{W})$, $j \in \mathcal{P}$ as a difference between two concave functions, where
\begin{equation}
\label{eqn5}
G_{j}(\mathbf{P},\boldsymbol{W}) = Q_{j}(\mathbf{P},\boldsymbol{W}) - S_{j}(\mathbf{P},\boldsymbol{W}),
\end{equation}
\vspace{-0.4cm}
\begin{align}
\label{eqn_2_x}
Q_{j}(\mathbf{P},\boldsymbol{W}) &= \text{log}_2 \biggl(\sum_{\substack{X\in\mathcal{U} \\ X\neq A_j}}  P_{X} \text{Tr}(\mathbf{\tilde{H}}_{X,A_j} \boldsymbol{W})  + \sigma^2 + \sigma_{l}^2\biggr) \nonumber \\&+ \text{log}_2 \biggl(\sum_{\substack{X\in\mathcal{U} \\ X\neq B_j}}  P_{X} \text{Tr}(\mathbf{\tilde{H}}_{X,B_j} \boldsymbol{W})  + \sigma^2 + \sigma_{l}^2\biggr) \nonumber \\&+ \text{log}_2 \biggl( \sum_{\substack{X\in\mathcal{U} \\ X\neq A_j,B_j}} P_{X}\text{Tr}(\mathbf{\tilde{H}}_{X,E}\boldsymbol{W})  + \sigma^2\biggr),
\end{align}
and
\begin{align}\label{eqn_2_y}
S_{j}(\mathbf{P},\boldsymbol{W}) &= \text{log}_2 \biggl(\sum_{\substack{X\in\mathcal{U} \nonumber \\ X\neq A_j,B_j}}  P_{X}\text{Tr}(\mathbf{\tilde{H}}_{X,A_j}\boldsymbol{W})  + \sigma^2 + \sigma_{l}^2\biggr) \\&+ \text{log}_2 \biggl(\sum_{\substack{X\in\mathcal{U} \nonumber \\ X\neq A_j,B_j}}  P_{X}\text{Tr}(\mathbf{\tilde{H}}_{X,B_j}\boldsymbol{W})  + \sigma^2 + \sigma_{l}^2\biggr) \\&+\text{log}_2 \biggl(\sum_{X\in\mathcal{U}} P_{X}\text{Tr}(\mathbf{\tilde{H}}_{X,E}\boldsymbol{W})  + \sigma^2\biggr).
\end{align}
In summary, the problem in epigraph form is expressed as
\begin{changetheequation}{P2}
\begin{subequations_opti}\label{eq:8}
\begin{align}
\text{(P2):} \quad \mathop{\max_{\boldsymbol{W},\mathbf{P}}}  & \quad  t \label{eq:8A}\\
\textrm{s.t.} \quad & G_{j}(\mathbf{P},\boldsymbol{W}) \geq t, & \quad & \forall j\in\mathcal{P},\label{eq:8B}\\
&\text{diag}(\boldsymbol{W}) = 1, \label{eq:8C}\\
  &\boldsymbol{W} \succeq 0,   \label{eq:8D}\\&\text{Rank}(\boldsymbol{W}) = 1 ,   \label{eq:8E}\\ & P_{\text{min}} \leq P_{X} \leq P_{\text{max}}, & \quad & \forall X\in\mathcal{U}. \label{eq:8F}
\end{align}
\end{subequations_opti}
\end{changetheequation}
Application of SDR requires constraining the feasible region of the variable $\boldsymbol{W}$ to positive semi-definite matrices, depicted by \eqref{eq:8D}. The diagonal constraint \eqref{eq:8C} stems from \eqref{eq:6B} and \eqref{eq:6C} in \eqref{eq:6}. The rank constraint \eqref{eq:8E} ensures that $\tilde{\boldsymbol{\omega}}$ can be recovered as $\boldsymbol{W} = \tilde{\boldsymbol{\omega}} \tilde{\boldsymbol{\omega}}^\dag$ at the end of optimization. Due to the coupled optimization variables and the non-convex constraint \eqref{eq:8B}, \eqref{eq:8} is still non-convex with respect to $\mathbf{P}$ and $\boldsymbol{W}$. To tackle this, we iteratively optimize $\mathbf{P}$ and $\boldsymbol{W}$ in an alternating manner.

Due to the difference between two concave functions in \eqref{eqn5}, we use SCA to optimize IRS phase shifts when the transmit powers are fixed. Around any feasible point $\hat{\boldsymbol{W}}$, an upper-bound of the concave function \eqref{eqn_2_y} can be obtained by using the first-order Taylor approximation \cite{yu2020robust} given by
\begin{equation}
\label{eqn7}
\begin{split}
\! S_{j}(\mathbf{P},\boldsymbol{W})\! \!  \leq \! \!  S_{j}(\mathbf{P},\hat{\boldsymbol{W}})\! + \!\text{Tr}\!\left(\!\nabla_{\boldsymbol{W}}^\dag S_{j}(\mathbf{P},\hat{\boldsymbol{W}})(\boldsymbol{W} \! -\! \hat{\boldsymbol{W}})\! \! \right),   
\end{split}
\end{equation}
where
\begin{equation}
\label{eqn_2_z_1}
\begin{split}
&\nabla_{\boldsymbol{W}} S_{j}(\mathbf{P},\boldsymbol{W}) \\&= \frac{1}{\text{ln}(2)} \Biggl\{ \frac{\sum_{\substack{X\in\mathcal{U} \\ X\neq A_j,B_j}}  P_{X}\mathbf{\tilde{H}}_{X,A_j}}{ \sum_{\substack{X\in\mathcal{U} \\ X\neq A_j,B_j}}  P_{X}\text{Tr}(\mathbf{\tilde{H}}_{X,A_j}\boldsymbol{W})  + \sigma^2 + \sigma_{l}^2}  \\&+ \frac{\sum_{\substack{X\in\mathcal{U} \\ X\neq A_j,B_j}}  P_{X}\mathbf{\tilde{H}}_{X,A_j}}{ \sum_{\substack{X\in\mathcal{U} \\ X\neq A_j,B_j}}  P_{X}\text{Tr}(\mathbf{\tilde{H}}_{X,A_j}\boldsymbol{W})  + \sigma^2 + \sigma_{l}^2} \\& +\frac{\sum_{X\in\mathcal{U}} P_{X}\mathbf{\tilde{H}}_{X,E}}{ \sum_{X\in\mathcal{U}} P_{X}\text{Tr}(\mathbf{\tilde{H}}_{X,E}\boldsymbol{W})  + \sigma^2} \Biggr\}
\end{split}
\end{equation}
and $\nabla_{\boldsymbol{W}}(.)$ represents the gradient with respect to $\boldsymbol{W}$. We substitute the upper-bound \eqref{eqn7} in \eqref{eq:8B} and drop the non-convex rank constraint \eqref{eq:8E}.
When $\mathbf{P}$ is fixed and known, we obtain the semi-definite programming (SDP) problem given by

\vspace{-0.4cm}

\begin{changetheequation}{P3}
\begin{subequations_opti}\label{eq:10}
\begin{align}
\text{(P3):} \enspace  \mathop{\max_{\boldsymbol{W},t}}  & \quad  t \label{eq:10A}\\
&\textrm{s.t.} \enspace  Q_{j}(\mathbf{P},\boldsymbol{W})  - S_{j}(\mathbf{P},\hat{\boldsymbol{W}})  \nonumber\\&  - \text{Tr}\left(\nabla_{\boldsymbol{W}}^\dag S_{j}(\mathbf{P},\hat{\boldsymbol{W}})(\boldsymbol{W}-\hat{\boldsymbol{W}})\right) \geq t,    \nonumber\\& \forall j\in\mathcal{P},\label{eq:10B}\\
&\text{diag}(\boldsymbol{W}) = 1, \label{eq:10C}\\
  &\boldsymbol{W} \succeq 0.   \label{eq:10D}\\
  &||\boldsymbol{W}-\hat{\boldsymbol{W}}|| \leq \xi.   \label{eq:10E}
\end{align}
\end{subequations_opti}
\end{changetheequation}
Using standard convex program solvers such as the CVX toolbox~\cite{cvx}, \eqref{eq:10} can be efficiently solved. By substituting $\hat{\boldsymbol{W}}$ with optimized $\boldsymbol{W}^\star$ at the end of each sub-iteration, IRS phases shifts are iteratively refined.
The constraint \eqref{eq:10E} limits the error of the approximation in \eqref{eqn7} \cite{wijewardena2021} and improves the convergence of the iterative IRS phase shift optimization.

\subsection{Transmit Power Optimization}

In this step, transmit powers are optimized with the IRS phase shifts fixed and known. We transform \eqref{eq:6} into epigraph form given by 
\vspace{-0.25cm}
\begin{changetheequation}{P4}
\begin{subequations_opti}\label{eq:9}
\begin{align}
\text{(P4):} \enspace \mathop{\max_{\mathbf{P},t}}  & \quad  t \label{eq:9A}\\
\textrm{s.t.} \quad & C_j(\mathbf{P},\boldsymbol{\omega})  \geq t, ~ j\in\mathcal{P},\label{eq:9B}\\& 
 P_{\text{min}} \leq P_{X} \leq P_{\text{max}}, ~ \forall X\in\mathcal{U}. \label{eq:9F}
\end{align}
\end{subequations_opti}
\end{changetheequation}
When $N = 1$, interference terms in \eqref{eqn_dbl_x} and \eqref{eqn_dbl_y} are absent. Consequently, \eqref{eq:9} converts into a simple maximization problem, where solving a convex optimization problem is not necessary, due to the monotonic properties of the objective function \cite{wijewardena2021}. In this case, the search space of \eqref{eq:9} is limited to corner points of the feasible region, which is not applicable for $N > 1$ due to the min function. Therefore, we treat \eqref{eq:9} as a fractional programming (FP) problem and apply quadratic transform technique to handle the SINR terms, following the direct FP method outlined in \cite{shen2018fractional}. Firstly, we define additional variables $x_{1,j}$, $x_{2,j}$, and $x_{3,j}$ for $j \in \mathcal{P}$ given by
\begin{equation}
\label{eqn6_1_x}
\begin{split}
 x_{1,j}(\mathbf{P}) =  \frac{\sqrt{P_{B_j}|\boldsymbol{\omega}^\dag \mathbf{H}_{B_j,A_j}|^2}}{\sum_{\substack{X\in\mathcal{U} \\ X\neq A_j,B_j}}  P_{X}|\boldsymbol{\omega}^\dag \mathbf{H}_{X,A_j}|^2 + \sigma^2 + \sigma_{l}^2},
\end{split}
\end{equation}
\begin{equation}
\label{eqn6_2_x}
\begin{split}
 x_{2,j}(\mathbf{P}) =  \frac{\sqrt{P_{A_j}|\boldsymbol{\omega}^\dag \mathbf{H}_{A_j,B_j}|^2}}{\sum_{\substack{X\in\mathcal{U} \\ X\neq A_j,B_j}}  P_{X}|\boldsymbol{\omega}^\dag \mathbf{H}_{X,B_j}|^2 + \sigma^2 + \sigma_{l}^2},
\end{split}
\end{equation}
and
\begin{equation}
\label{eqn6_3_x}
\begin{split}
 x_{3,j}(\mathbf{P}) =  \frac{\sqrt{P_{A_j}|\boldsymbol{\omega}^\dag \mathbf{H}_{A_j,E}|^2 + P_{B_j}|\boldsymbol{\omega}^\dag \mathbf{H}_{B_j,E}|^2}}{\sum_{\substack{X\in\mathcal{U} \\ X\neq A_j,B_j}}  P_{X}|\boldsymbol{\omega}^\dag \mathbf{H}_{X,E}|^2 + \sigma^2}.
\end{split}
\end{equation}
We substitute \eqref{eqn6_1_x}, \eqref{eqn6_2_x}, and \eqref{eqn6_3_x} appropriately in \eqref{eqn3} to derive a function that is both equivalent to \eqref{eqn3} and convex with respect to $\mathbf{P}$, given by $f_j(\mathbf{P},\boldsymbol{\omega},\mathbf{x}_j) $ $ = \text{log}_2 \Bigl(\!1+ 2x_{1,j}\sqrt{P_{B_j}|\boldsymbol{\omega}^\dag \mathbf{H}_{B_j,A_j}|^2}  - x_{1,j}^2 \allowbreak \bigl( \allowbreak \sum_{\substack{X\in\mathcal{U} \\ X\neq A_j,B_j}}  P_{X}|\boldsymbol{\omega}^\dag \mathbf{H}_{X,A_j}|^2 + \sigma^2 + \sigma_{l}^2\bigl)\Bigl)$ $+ \text{log}_2 \Bigl(\!1+ 2x_{2,j} \allowbreak \sqrt{P_{A_j}\allowbreak|\boldsymbol{\omega}^\dag \mathbf{H}_{A_j,B_j}|^2}  - x_{2,j}^2\bigl(\sum_{\substack{X\in\mathcal{U} \\ X\neq A_j,B_j}}  P_{X}|\boldsymbol{\omega}^\dag \mathbf{H}_{X,B_j}|^2 + \sigma^2 + \sigma_{l}^2\bigr)\Bigr)$ $+\text{log}_2\Bigl(\!\!1\! - x_{3,j}^2\bigl(\sum_{\substack{X\in\mathcal{U} \\ X\neq A_j,B_j}}  P_{X}|\boldsymbol{\omega}^\dag \mathbf{H}_{X,E}|^2 +\!2x_{3,j}\sqrt{P_{A_j}|\boldsymbol{\omega}^\dag \mathbf{H}_{A_j,E}|^2 \!+\! P_{B_j}|\boldsymbol{\omega}^\dag \mathbf{H}_{B_j,E}|^2} + \sigma^2\bigr)\Bigr)$, where $\mathbf{x}_j = [x_{1,j}, x_{2,j}, x_{3,j}]$, $j \in \mathcal{P}$. Assuming $\mathbf{x}_j$, $j \in \mathcal{P}$ are computed using the initialized $\mathbf{P}$, \eqref{eq:9} can be expressed as
\begin{changetheequation}{P5}
\begin{subequations_opti}\label{eq:11}
\begin{align}
\text{(P5):} \enspace \mathop{\max_{\mathbf{P},t}}  & \quad  t \label{eq:11A}\\
\textrm{s.t.} \quad & f_j(\mathbf{P},\boldsymbol{\omega},\mathbf{x}_j)  \geq t, ~ j\in\mathcal{P},\label{eq:11B}\\& 
 P_{\text{min}} \leq P_{X} \leq P_{\text{max}}, ~ \forall X\in\mathcal{U}. \label{eq:11C}
\end{align}
\end{subequations_opti}
\end{changetheequation}
Problem \eqref{eq:11} is a convex problem, which can be efficiently solved using standard convex program solvers. Computation of $\mathbf{x}_j$, $j \in \mathcal{P}$ followed by evaluating \eqref{eq:11} are iterated until convergence.

\subsection{Alternating Optimization Algorithm}
In this step, we summarize the SCA and FP based AO algorithm. Algorithm \ref{alg:scpao} integrates the two sub-iterations of IRS phase shift optimization and power optimization, where each sub-iteration terminates when the fractional increase is below $\epsilon_1$ and $\epsilon_2$ respectively. The overall algorithm terminates when the minimum secrecy rate reaches an fractional increase, below $\epsilon_3$. 
At the end of the algorithm, $\hat{\boldsymbol{W}}$ is not necessarily rank-one. Therefore, we apply a rank-one operation such as Gaussian randomization \cite{luo2010semidefinite} to obtain a rank-one approximation $\boldsymbol{\hat{\omega}}$ of $\boldsymbol{\omega}$. The final estimate of $\boldsymbol{\omega}_{ind}$ is expressed as
\begin{equation}
\label{eqnrecoverphases}
\begin{split}
 \boldsymbol{\omega}_f = \text{exp}\left(j  \angle\left( \left[ \frac{\boldsymbol{\hat{\omega}}}{\hat{\omega}_{L+1}} \right]_{(1:L)} \right)\right)
\end{split}
\end{equation}
where $\angle(\boldsymbol{x})$ is denotes the phases of the complex vector $\boldsymbol{x}$. The transmit powers are obtained as $\boldsymbol{P}_f = \hat{\boldsymbol{P}}$.

\begin{algorithm}
\caption{SCA and FP based AO algorithm}\label{alg:scpao}
\begin{algorithmic}[1]

\State Initialize $\hat{\boldsymbol{W}} = \boldsymbol{\omega} \boldsymbol{\omega}^\dag$ such that $\boldsymbol{\omega} \in \mathbb{C}^{(L+1)}$ and $|\omega^{(i)}| = 1$ for $1 \leq i \leq L + 1$,  $\hat{\boldsymbol{P}}$ such that $P_{X} = P_{\text{max}}, ~ \forall X\in\mathcal{U}$
\Repeat
\Repeat
\State Compute $\mathbf{x}_j(\hat{\boldsymbol{P}})$, $j \in \mathcal{P}$ using \eqref{eqn6_1_x}, \eqref{eqn6_2_x}, \eqref{eqn6_3_x}
\State Fix $\boldsymbol{W} = \hat{\boldsymbol{W}}$ and solve \eqref{eq:11} to obtain $\boldsymbol{P}^{\star}$ \State Set $\hat{\boldsymbol{P}} = \boldsymbol{P}^{\star}$
\Until{Fractional increase in \eqref{eqn3} $\leq \epsilon_1$}
\Repeat
\State Fix $\boldsymbol{P} = \hat{\boldsymbol{P}}$ and solve \eqref{eq:10} to obtain $\boldsymbol{W}^{\star}$
\State  Set $\hat{\boldsymbol{W}} = \boldsymbol{W}^{\star}$
\Until{Fractional increase in \eqref{eqn3} $\leq \epsilon_2$}
\Until{Fractional increase in \eqref{eqn3} $\leq \epsilon_3$}
\State Estimate $\boldsymbol{\hat{\omega}} =\text{Rank\_one\_operation}(\hat{\boldsymbol{W}})$
\State Output $\boldsymbol{\omega}_f$ using \eqref{eqnrecoverphases} and $\boldsymbol{P}_f = \hat{\boldsymbol{P}}$
\end{algorithmic}
\end{algorithm} 

The time complexity of algorithm 1 is a function of both $L$ and $N$. IRS phase shift optimization uses semi-definite programming (SDP), whose complexity scales rapidly with the size of positive semi-definite (PSD) matrix. Since, the number of constraints in \eqref{eq:10} containing the PSD matrix $\boldsymbol{W}$ is $N + 2$, the complexity of \eqref{eq:10} is obtained as $\mathcal{O}\left(\sqrt{L} \left(N L^3+N^2 L^2+N^3 \right)\right)$ \cite{polik2010interior, yu2020robust}. With the total number of decision variables $2N$ and the total number of constraints $3N$, the complexity of \eqref{eq:11} in transmit power optimization is obtained as $\mathcal{O}\left(N^{3.5}\right)$ \cite{xia2024joint}.

\section{Numerical Results}

\begin{figure}[!t]
\centering
\includegraphics[width=\columnwidth]{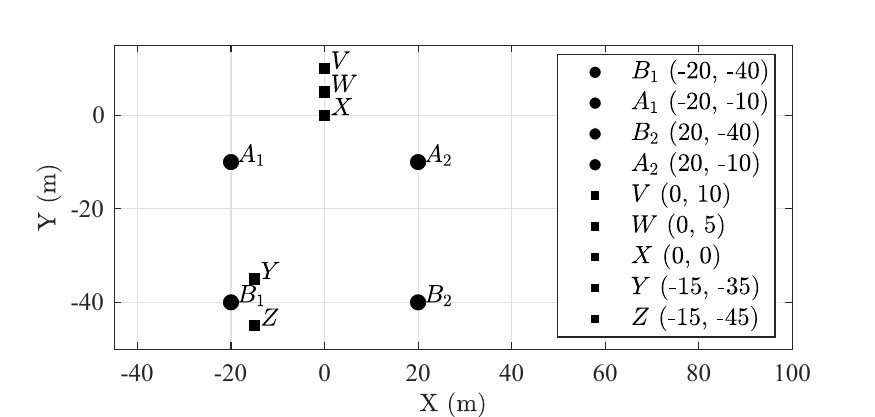}
\vspace{-0.4cm}
\caption{Positions of all nodes considered in the numerical results.}
\label{fig_scenarios}
\vspace{-0.2cm}
\end{figure}

\begin{figure}[!t]
\centering
\includegraphics[width=\columnwidth]{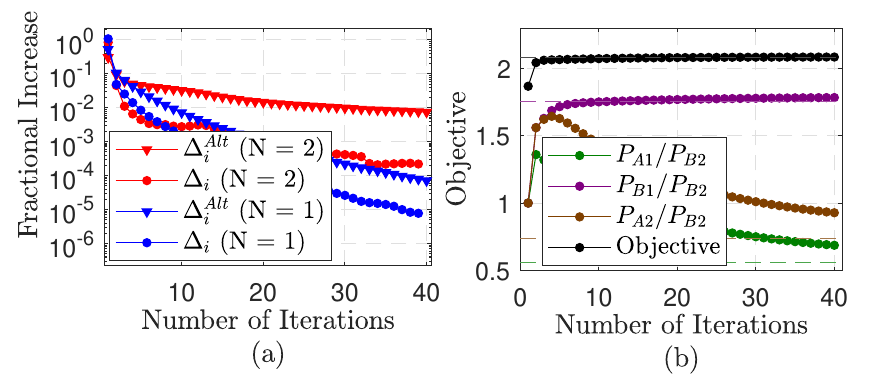}
\vspace{-0.6cm}
\caption{(a) Average fractional increase in minimum secrecy rate with respect to the number of sub-iterations in power optimization. (b) Variation of power optimization variables and the objective with increasing sub-iterations. Dashed lines indicate the values obtained using the alternative method.}
\label{fig_conv1}
\vspace{-0.6cm}
\end{figure}

\begin{figure}[!t]
\centering
\includegraphics[width=\columnwidth]{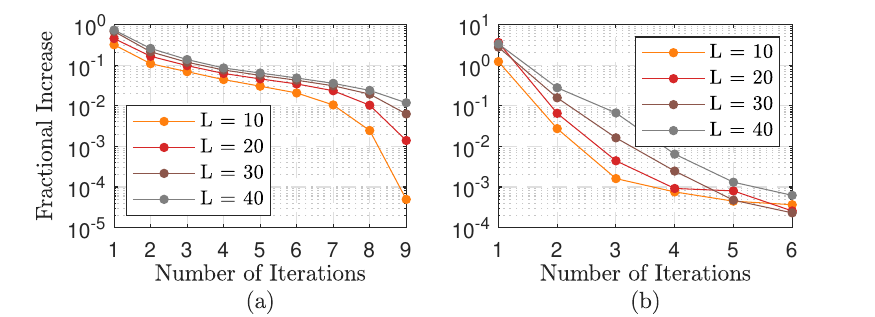}
\vspace{-0.4cm}
\caption{(a) Average fractional increase in minimum secrecy rate with respect to the number of sub-iterations, when IRS phase shifts are optimized. (b) Average fractional increase in minimum secrecy rate with respect to the number of outer-iterations in Algorithm 1.}
\label{fig_conv2}
\vspace{-0.4cm}
\end{figure}

In this section, we illustrate the performance of the proposed algorithm using the scenario of $N = 2$. Four users are assumed to be in fixed positions as shown in Fig. \ref{fig_scenarios}. Malicious user and the IRS are appropriately positioned at V, W, X, Y and Z to create four user configurations as shown in the Table \ref{table:1}. The channel between the IRS and an arbitrary user $A_j$ is modelled as $\boldsymbol{f}_j = \sqrt{L_0 d^{-\alpha}} \left( \sqrt{\frac{\beta}{\beta + 1}}\boldsymbol{f}_j^{LoS} + \sqrt{\frac{1}{\beta + 1}}\boldsymbol{f}_j^{NLoS} \right)$,
where $L_0$, $d$, $\alpha$ and $\beta$ denote the path loss at reference distance $1$\,m, the distance between the two nodes, the path loss exponent, and the Rician $K$-factor, respectively. $\boldsymbol{f}_j^{LoS}$ and $\boldsymbol{f}_j^{NLoS}$ denote the line-of-sight (LoS) and non-LoS components of the channel. All other channels are modeled in a similar manner. Following channel parameters are considered in the simulation. The user-IRS channels and the user-user channels are subject to path-loss exponents $\alpha = 2$ and $\alpha = 3$, respectively. The user-IRS channels are modeled as Rician fading channels with $\beta = 8$, whereas the user-user channels are modeled as Rayleigh fading channels, corresponding to $\beta = 0$. The transmit power limits of each user are assumed to be $P_{\text{min}} = 0\,\text{dBm}$ and $P_{\text{max}} = 15\,\text{dBm}$. We set $L_0 = -30\,\text{dB}$, $\sigma^2 = -105\,\text{dBm}$ and $\sigma_{l}^2 = -100\,\text{dBm}$. The values of $\epsilon_1$, $\epsilon_2$ and $\epsilon_3$ in the algorithm are set to $10^{-2}$. The convergence results in Fig. \ref{fig_conv1} and Fig. \ref{fig_conv2} are generated using the user configuration C1.
\begin{table}[!t]
\centering
\caption{Locations of E and the IRS in different configurations.}
\label{table:1}
\begin{tabular}{|c|c|c|}
\hline
\textbf{User configuration and description} & \textbf{E} & \textbf{IRS} \\
\hline
C1-IRS is close to B1. E is far from users. & W & Z \\
\hline
C2-IRS and E, 5m apart, away from users. & W & V \\
\hline
C3-E is close to B1. IRS is far from users. & Z & V \\
\hline
C4-Both E and IRS are close to B1. & Y & Z \\
\hline
\end{tabular}
\vspace{-0.4cm}
\end{table}

With the aim of assessing the convergence of our power optimization method for $N \geq 2$, we compare the FP method with an alternative baseline from two perspectives. First we compare how the value of the optimization objective evolves over iterations. Then, we compare how the values of the variables evolve in the search space. The fractional increase in the objective measured relative to its value in the previous iteration is given by $\Delta_i \triangleq \frac{J^{\text{FP}}_i - J^{\text{FP}}_{i-1}}{J^{\text{FP}}_{i-1}}$,
where $J^{\text{FP}}_i$ denotes the value of the objective in $i^\text{th}$ sub-iteration, when optimized using the FP method. Fig. \ref{fig_conv1}(a) includes the variation of $\Delta_i$ with the number of sub-iterations averaged over channel realizations, when $L = 20$ and the IRS phase shifts are fixed at random values. Convergence is faster for $N = 1$, compared to $N = 2$.
Furthermore, we define $\Delta_i^\text{Alt} \triangleq \frac{ J^{\text{Alt}}-J^{\text{FP}}_i}{J^{\text{Alt}}}$,
where $J^{\text{Alt}}$ denotes the objective optimized using an alternative method. For N = 1, authors of \cite{wijewardena2021} propose an optimal alternative optimization method, where the optimization is performed among four corner points of the feasible region. For $N = 2$, the alternative method is a grid search, where each optimization variable is assigned $Q$ evenly spaced values across its feasible range resulting in $Q^4$ evaluations of the objective function. In this paper, $Q$ is set to 50. When $N = 1$, FP method demonstrates to be a good choice for power optimization since $\Delta_i^\text{Alt}$ approaches $10^{-2}$ within 10 iterations. In contrast, when $N = 2$, it requires 30 iterations for $\Delta_i^\text{Alt}$ to reach $10^{-2}$, due to the added constraints from the max-min optimization and increased number of optimization variables. Fig. \ref{fig_conv1}(b) shows the behaviour of optimization variables and the objective with increasing number of sub-iterations, when the optimization is performed for a specific channel realization for $N = 2$. For comparison, same parameters obtained using the grid search are shown using dashed lines. Observe that optimization variables represented as ratios approach the values obtained using exhaustive method with increasing iteration number. Therefore, convergence is achieved for $N>1$, albeit requiring more sub-iterations.

Fig. \ref{fig_conv2}(a) shows the average fractional increase in minimum secrecy rate with the number of sub-iterations, when IRS phase shifts are optimized with the transmit powers fixed at $\boldsymbol{P} = \hat{\boldsymbol{P}}$. It can be seen that the number of iterations required to reach a specific fractional increase value increases with $L$. Average fractional increase in minimum secrecy rate with the number of outer-iterations in Algorithm 1 is shown in Fig. \ref{fig_conv2}(b). Observe that within 4 outer-iterations, Algorithm 1 achieves a fractional increase below $10^{-2}$. In this paper, Algorithm 1 is implemented such that each outer iteration consists of 25 power optimization sub-iterations and 8 IRS phase shift optimization sub-iterations. We use 4 outer iterations for a given channel realization.

\begin{figure}[!t]
\centering
\includegraphics[width=\columnwidth]{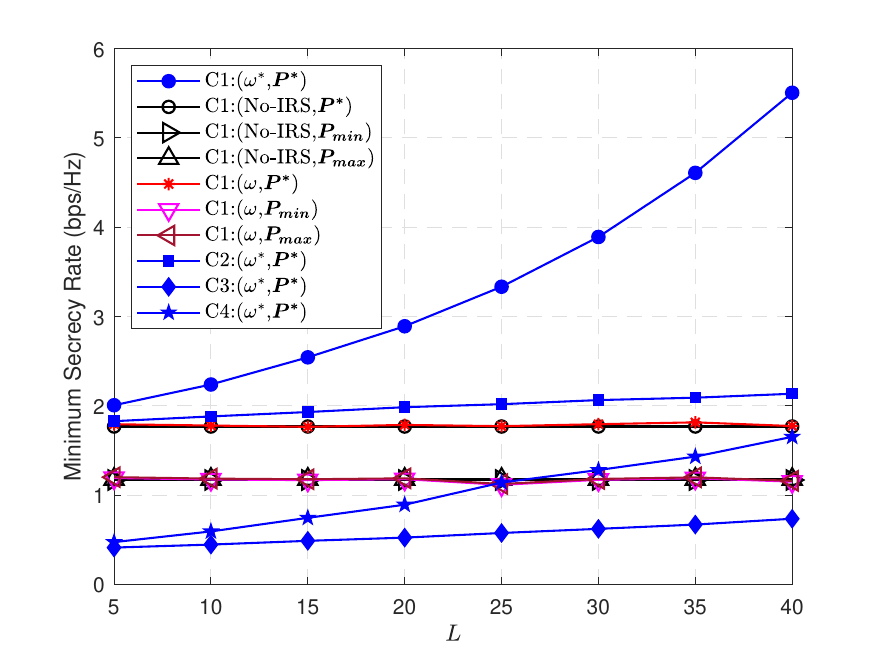}
\vspace{-0.4cm}
\caption{Variation of minimum secrecy rate with the number of IRS elements in four different user configurations. For configuration C1, six scenarios are created using no-IRS, IRS with random phase shifts, minimum power, maximum power, and optimized power conditions.}
\label{fig_configs}
\vspace{-0.4cm}
\end{figure}

Fig. \ref{fig_configs} illustrates the improvement in the minimum secrecy rate as the number of IRS elements increases. All four configurations in the Table \ref{table:1} are considered. Among the configurations, C1 exhibits a high minimum secrecy rate for all $L$, and fast growth in minimum secrecy rate with increasing $L$. With close proximity to $B_1$, IRS improves the information rate among the user pairs and suppresses inter-user interference with increasing $L$. With the malicious user residing far away from users, information leakage is minimal as expected. It is interesting to notice, $\text{C1:}(\omega,P_\text{min})$, $\text{C1:}(\omega,P_\text{max})$, $\text{C1:}(\text{No-IRS},P_\text{min})$ and $\text{C1:}(\text{No-IRS},P_\text{max})$ overlap in performance. Furthermore, $\text{C1:}(\omega,P^\star)$ and $\text{C1:}(\text{No-IRS},P^\star)$ overlap in performance. Thus, having an IRS with random phase shifts is equivalent to having no IRS. Setting all transmit powers to maximum or minimum values results in identical sub-optimal performance. Compared to the baseline scheme $\text{C1:}(\omega,P^\star)$, C1 achieves a gain of approximately 211\% with $L = 40$. Similarly, C2, C3 and C4 achieve gains reaching 22\%, 92\% and 359\%, respectively. This gain is desired since the minimum secrecy rate reflects the weakest user pair.

For small values of $L$, C2 exhibits similar performance as C1, although the growth with increasing $L$ is negligible compared to C1. When the IRS is far away from users, the ability of the IRS to assist the information rate of legitimate users is low. In \eqref{eqn_dbl_y}, the terms in the numerator contributes to information leakage, whereas the terms in the denominator are interference terms that inhibits the information leakage. Reducing the information leakage for first user pair increases the information leakage for second user pair, since the numerator terms in $I(y_E;s_{A_1},s_{B_1})$ are available in the denominator of $I(y_E;s_{A_2},s_{B_2})$, and vice versa. Therefore, even when the malicious user is close to the IRS, capability of the IRS to reduce information leakage while maintaining the fairness is limited. C3 exhibits the worst performance, since the information leakage to the closely residing malicious user is very high and it is difficult for the IRS to assist the information rate of legitimate users without close proximity to users. Note that the first user pair is highly disadvantaged, and minimum secrecy rate reflects its performance. In contrast, both the IRS and the malicious user are close to $B_1$ in C4. Due to the IRS's proximity to disadvantaged $B_1$, the worst performance of C3 is improved in C4.


\vspace{-0.1cm}

\section{Conclusions}

This paper has investigated an algorithm for optimizing an IRS assisted multi-user secure two-way communication system. Transmit powers and IRS phase shifts have been iteratively optimized to maximize the minimum secrecy rate among user pairs. The convergence of the proposed algorithm has been evaluated. The performance gain in minimum secrecy rate among user pairs has been obtained for four different configurations. Results show that user fairness improves the most, when the IRS is positioned near the legitimate users.

\bibliographystyle{IEEEtran}
\bibliography{IEEEabrv,IEEEexample}

\begin{thebibliography}{10}
\providecommand{\url}[1]{#1}
\csname url@samestyle\endcsname
\providecommand{\newblock}{\relax}
\providecommand{\bibinfo}[2]{#2}
\providecommand{\BIBentrySTDinterwordspacing}{\spaceskip=0pt\relax}
\providecommand{\BIBentryALTinterwordstretchfactor}{4}
\providecommand{\BIBentryALTinterwordspacing}{\spaceskip=\fontdimen2\font plus
\BIBentryALTinterwordstretchfactor\fontdimen3\font minus \fontdimen4\font\relax}
\providecommand{\BIBforeignlanguage}[2]{{%
\expandafter\ifx\csname l@#1\endcsname\relax
\typeout{** WARNING: IEEEtran.bst: No hyphenation pattern has been}%
\typeout{** loaded for the language `#1'. Using the pattern for}%
\typeout{** the default language instead.}%
\else
\language=\csname l@#1\endcsname
\fi
#2}}
\providecommand{\BIBdecl}{\relax}
\BIBdecl

\bibitem{wu2019intelligent}
Q.~Wu and R.~Zhang, ``Intelligent reflecting surface enhanced wireless network via joint active and passive beamforming,'' \emph{{IEEE} Trans. Wireless Commun.}, vol.~18, no.~11, pp. 5394--5409, 2019.

\bibitem{cui2019secure}
M.~Cui, G.~Zhang, and R.~Zhang, ``Secure wireless communication via intelligent reflecting surface,'' \emph{{IEEE} Wireless Commun. Lett.}, vol.~8, no.~5, pp. 1410--1414, 2019.

\bibitem{yu2019enabling}
X.~Yu, D.~Xu, and R.~Schober, ``Enabling secure wireless communications via intelligent reflecting surfaces,'' in \emph{Proc. IEEE Glob. Commun. Conf. (GLOBECOM)}, 2019, pp. 1--6.

\bibitem{yu2020robust}
X.~Yu, D.~Xu, Y.~Sun, D.~W.~K. Ng, and R.~Schober, ``Robust and secure wireless communications via intelligent reflecting surfaces,'' \emph{{IEEE} J. Sel. Areas Commun.}, vol.~38, no.~11, pp. 2637--2652, 2020.

\bibitem{xu2019resource}
D.~Xu, X.~Yu, Y.~Sun, D.~W.~K. Ng, and R.~Schober, ``Resource allocation for secure {IRS}-assisted multiuser {MISO} systems,'' in \emph{Proc. IEEE Globecom Workshops (GC Wkshps)}, 2019, pp. 1--6.

\bibitem{hong2020artificial}
S.~Hong, C.~Pan, H.~Ren, K.~Wang, and A.~Nallanathan, ``Artificial-noise-aided secure {MIMO} wireless communications via intelligent reflecting surface,'' \emph{{IEEE} Trans. Commun.}, vol.~68, no.~12, pp. 7851--7866, 2020.

\bibitem{dong2020secure}
L.~Dong and H.-M. Wang, ``Secure {MIMO} transmission via intelligent reflecting surface,'' \emph{{IEEE} Commun. Lett.}, vol.~9, no.~6, pp. 787--790, 2020.

\bibitem{chen2019}
J.~Chen, Y.-C. Liang, Y.~Pei, and H.~Guo, ``Intelligent reflecting surface: A programmable wireless environment for physical layer security,'' \emph{{IEEE} Access}, vol.~7, pp. 82\,599--82\,612, 2019.

\bibitem{atapattu2020}
S.~Atapattu, R.~Fan, P.~Dharmawansa, G.~Wang, J.~Evans, and T.~A. Tsiftsis, ``Reconfigurable intelligent surface assisted two--way communications: Performance analysis and optimization,'' \emph{{IEEE} Trans. Commun.}, vol.~68, no.~10, pp. 6552--6567, 2020.

\bibitem{wijewardena2021}
M.~Wijewardena, T.~Samarasinghe, K.~T. Hemachandra, S.~Atapattu, and J.~S. Evans, ``Physical layer security for intelligent reflecting surface assisted two--way communications,'' \emph{{IEEE} Commun. Lett.}, vol.~25, no.~7, pp. 2156--2160, 2021.

\bibitem{lv2020secure}
L.~Lv, Q.~Wu, Z.~Li, N.~Al-Dhahir, and J.~Chen, ``Secure two-way communications via intelligent reflecting surfaces,'' \emph{{IEEE} Commun. Lett.}, vol.~25, no.~3, pp. 744--748, 2020.

\bibitem{wang2012distributed}
H.-M. Wang, Q.~Yin, and X.-G. Xia, ``Distributed beamforming for physical-layer security of two-way relay networks,'' \emph{{IEEE} Trans. Signal Process.}, vol.~60, no.~7, pp. 3532--3545, 2012.

\bibitem{cvx}
M.~Grant and S.~Boyd, ``{CVX}: Matlab software for disciplined convex programming,'' 2008.

\bibitem{shen2018fractional}
K.~Shen and W.~Yu, ``{Fractional programming for communication systems—Part I: Power control and beamforming},'' \emph{{IEEE} Trans. Signal Process.}, vol.~66, no.~10, pp. 2616--2630, 2018.

\bibitem{luo2010semidefinite}
Z.-Q. Luo, W.-K. Ma, A.~M.-C. So, Y.~Ye, and S.~Zhang, ``Semidefinite relaxation of quadratic optimization problems,'' \emph{{IEEE} Signal Process. Mag.}, vol.~27, no.~3, pp. 20--34, 2010.

\bibitem{polik2010interior}
I.~P{\'o}lik and T.~Terlaky, ``Interior point methods for nonlinear optimization,'' in \emph{Nonlinear Optimization}.\hskip 1em plus 0.5em minus 0.4em\relax Springer, Jul. 2010, pp. 215--276.

\bibitem{xia2024joint}
F.~Xia, Z.~Fei, X.~Wang, P.~Liu, J.~Guo, and Q.~Wu, ``{Joint Waveform and Reflection Design for Sensing-Assisted Secure RIS-Based Backscatter Communication},'' \emph{{IEEE} Wireless Commun. Lett.}, 2024.

\end{thebibliography}

\end{document}